\documentclass[aps,prl,reprint,twocolumn,groupedaddress,amsmath,amssymb]{revtex4-1}
\usepackage{graphicx}  
\usepackage{dcolumn}   
\usepackage{bm}        
\usepackage{verbatim}   

\usepackage[pscoord]{eso-pic}
\newcommand{\placetextbox}[3]{
  \setbox0=\hbox{#3}
  \AddToShipoutPictureFG*{
    \put(\LenToUnit{#1\paperwidth},\LenToUnit{#2\paperheight}){\vtop{{\null}\makebox[0pt][c]{#3}}}%
  }%
}%
\usepackage{color}
\newcommand{\updated}{}

\usepackage{xspace}
\usepackage{amsmath}

\newcommand{\mt}{\ensuremath{{m}_{{t}}}\xspace}
\newcommand{\mtpole}{\ensuremath{m_t^{\text{p}}}\xspace}
\newcommand{\mtmc}{\ensuremath{m_t^{\text{MC}}}\xspace}

\newcommand{\mtmsbar}{\ensuremath{\overline{m}_t}\xspace}

\newcommand{\mtpoleconv}{\ensuremath{m_t^\text{p,c}}\xspace}
\newcommand{\GeV}{\ensuremath{\,\text{Ge\hspace{-.08em}V}}\xspace}
\newcommand{\TeV}{\ensuremath{\,\text{Te\hspace{-.08em}V}}\xspace}

\newcommand{\ttbar}    {\ensuremath{\text{t}\bar{\text{t}}}\xspace}

\newcommand{\xsec}    {\ensuremath{\sigma}\xspace}

\newcommand{\mlb}    {\ensuremath{m_{lb}}\xspace}
\newcommand{\MSbar}{\ensuremath{\overline{{\text{MS}}}}\xspace}
\newcommand{\alphas}{\ensuremath{\alpha_S}\xspace}

\newcommand{\obs}{\ensuremath{\xi}\xspace}

\newcommand{\lbd}{\ensuremath{\vec{\lambda}}\xspace}

\begin{document}

\title{Calibration of the Top-Quark Monte-Carlo Mass}
\author{Jan Kieseler}
\affiliation{
   Deutsches Elektronen Synchrotron DESY,
   Notkestr. 85,
   D-22607 Hamburg, Germany
   }
   \author{Sven-Olaf Moch}
   \affiliation{II. Institut f\"ur Theoretische Physik, 
   Universit\"at Hamburg,
   Luruper Chaussee 149,
   D-22761 Hamburg, Germany}
   \author{Katerina Lipka}
\affiliation{
   Deutsches Elektronen Synchrotron DESY,
   Notkestr. 85,
   D-22607 Hamburg, Germany
   }
\date{\today}
\begin{abstract}
We present a method to establish experimentally the relation between the top-quark mass
\mtmc as implemented in Monte-Carlo generators and the Lagrangian mass parameter \mt in a 
theoretically well-defined renormalization scheme.
We propose a simultaneous fit of \mtmc and an observable sensitive to \mt, 
which does not rely on any prior assumptions about the relation between \mt and \mtmc.
The measured observable is independent of \mtmc and can be used subsequently for a determination of \mt. 
The analysis strategy is illustrated with examples for the extraction of \mt 
from inclusive and differential cross sections for hadro-production of top-quarks.
\end{abstract}

\placetextbox{0.144}{0.97}{\textbf {DESY 15-190}}

\maketitle
\section{Introduction}
The top-quark mass is one of the fundamental parameters of the Standard Model
(SM). Its value significantly affects predictions for many observables either
directly or via radiative corrections. As a consequence, the measured
top-quark mass is one of the crucial inputs to electroweak precision fits,
which enable comparisons between experimental results and predictions within
and beyond the SM~\cite{Agashe:2014kda}. 
Furthermore, together with the Higgs-boson mass, it has
critical implications on the stability of the electroweak
vacuum~\cite{Bezrukov:2012sa,Degrassi:2012ry,Alekhin:2012py}. 

In fixed-order and analytically resummed predictions, the top-quark mass appears as a
parameter of the Lagrangian and, therefore, depends on the choice of the renormalization scheme once
corrections beyond leading order (LO) are consistently included.  
The conventional scheme choice in many applications of Quantum Chromodynamics (QCD) is the pole mass \mtpole, 
while alternative definitions based on the (modified) minimal subtraction realize the concept of a running mass
$\mtmsbar(\mu)$ at a renormalization scale $\mu$ as a particular example of so-called short-distance masses.
On the other hand, Monte Carlo (MC) simulations generally contain not only hard-interaction calculations at LO or
next-to leading order (NLO), with the fixed-order matrix elements as functions
of the top-quark's pole mass \mtpole,
but also contributions from initial and final
state radiation, hadronization, as well as underlying-event interactions,
modeled by parton shower programs based on leading-logarithm approximations
and heuristic models.
\updated{All these effects can lead to systematic shifts in the value of the top-quark mass~\cite{Skands:2007zg}.} 
Therefore, \updated{MC simulations presently} do not allow for a precise definition of the quark mass renormalization scheme. 

The top-quark mass has been determined with remarkable precision: the current
world average quoted as 173.34 $\pm$ 0.76\GeV is obtained by combining  results from
the Tevatron and the LHC~\cite{ATLAS:2014wva}.  
However, these measurements rely on the relation between the top-quark mass
and the respective experimental observable, e.g., the reconstructed invariant
mass of the top-quark decay products. 
This relation is derived by using MC simulations, 
so that these measurements determine the top-quark mass parameter implemented in these simulations.
Therefore, the determined parameter is the so-called Monte-Carlo mass \mtmc, 
which appears most appropriate to describe experimental data~\cite{Agashe:2014kda,ATLAS:2014wva,Buckley:2011ms}.

The unambiguous interpretation of the experimental results for \mtmc in terms of 
a Lagrangian top-quark mass \updated{($\mt$)} in a specific renormalization scheme employed
in the SM has been a longstanding and increasingly urgent problem, 
given the importance of the value of the top-quark mass for SM physics analysis and 
the small uncertainty in the experimental measurement of \mtmc~\cite{ATLAS:2014wva}. 
At present, the translation from \mtmc to a theoretically well-defined mass definition 
in a short-distance scheme at a low scale can only be estimated 
to be ${\cal O}(1)$~GeV, see, e.g., Ref.~\cite{Hoang:2008xm,Moch:2014tta}.

In consequence, a measurement of \mt is preferable and can be performed by
confronting a measured observable sensitive to \mt with its prediction,
calculated at NLO in QCD or beyond in a well-defined renormalization scheme for the top-quark mass. 
For this purpose, the inclusive \updated{cross section ($\xsec$)} and the normalized differential \updated{cross sections for 
top-quark pair ({\ttbar}) production} have been employed to determine the pole mass~\cite{CMS-PAS-TOP-13-004,Aad:2014kva,Aad:2015waa}.
For these measurements of \mtpole, detector and process modeling effects are
evaluated using MC simulations, so that the measured observable typically depends on \mtmc. 
\updated{Even though the extracted value of \mtpole does not depend on a specific \mtmc hypothesis, 
it relies on the relation between both parameters, the exact difference ($\Delta_m^p=\mtpole-\mtmc$) 
being unknown. However, it is often assumed to be up to 1\GeV, leading to a systematic uncertainty on the measurement~\cite{CMS-PAS-TOP-13-004,Aad:2014kva,Aad:2015waa}, which might be under- or overestimated.}
This uncertainty can be \updated{small} when only the shape of a particular observable
 defined within the detectors fiducial volume is considered~\cite{Aad:2015waa}, 
since the dependence on \mtmc mainly enters through detector-acceptance effects. 
However, the sensitivity to \mt increases when the total \ttbar production rate is also taken into account.

The pole mass scheme, which is inspired by the definition of the electron mass 
in Quantum Electrodynamics, has short-comings when applied to quarks
in a confined theory~\cite{Bigi:1994em,Beneke:1994sw}. 
Non-perturbative corrections to \mtpole due to the infrared renormalon 
lead to an intrinsic theoretical ambiguity 
of the order of $\Lambda_\text{QCD}$~\cite{Bigi:1994em,Beneke:1994sw,Smith:1996xz}. 
Alternatively, \xsec can be calculated using other mass 
schemes~\cite{Hoang:1998hm,Hoang:1999zc,Beneke:1998rk,Langenfeld:2009wd}, such as
the aforementioned running mass definition at a scale $\mu$, $\mtmsbar(\mu)$, 
the so-called \MSbar mass.
By using \mtmsbar in the calculation of \xsec, the perturbative expansion in
the strong coupling exhibits a significantly faster
convergence~\cite{Langenfeld:2009wd}. 

This letter describes a generic approach to measure an observable $\obs$
sensitive to \mt in a particular renormalization scheme without any prior assumptions on \mtmc or its relation to \mt. 
The method employs a simultaneous likelihood fit of \mtmc and \obs, comparing
an observed distribution in data to its MC prediction. 
\updated{For the latter, two categories of processes are taken into account. The first one corresponds to the signal process, i.e. 
top-quark pair production or single top-quark production, for which the cross section and event kinematics depend on \mt. The 
second category comprises background processes such as e.g. the production of electroweak bosons and shows no significant 
dependence on \mt.}
Subsequently, a determination of \mt can be performed in a given renormalization scheme 
comparing data to theory predictions for $\obs(\mt)$ and, therefore, a calibration of \mtmc
by quantifying the difference $\Delta_m=\mt-\mtmc$ is possible.
The method is first discussed for the special case with \obs being an
inclusive signal production cross section and extended to differential cross
sections in a second step.

\section{Calibration with inclusive cross sections}

\updated{Assume, to measure the inclusive cross section \xsec, a number of {\it detected} events, $N^d$,} is reconstructed
and selected experimentally, with an \updated{\it efficiency} $\epsilon$ estimated by using simulation.
\updated{In total, $N^p$ expected events are confronted with those observed in data. We propose to perform this comparison in
bins of an observable sensitive to \mtmc.}
The parameterization is chosen such that the shape of the distribution constrains \mtmc, while its normalization determines \xsec. 
For this purpose, the fraction of predicted  signal events $n_i^p$ in bin $i$ 
is considered and the total number of predicted events $N^p_i$ in the same bin is written as: 
\begin{equation}
\label{eq:Npi}
N^p_i=\mathcal{L}\cdot  \epsilon(\mtmc,\lbd)\cdot \xsec \cdot  n_i^p(\mtmc,\lbd)  +N^{bg}_i(\lbd) \text{,}
\end{equation}
with $N^{bg}_i$ being the contribution from background processes and $\mathcal{L}$ the integrated luminosity. 
Systematic uncertainties due to detector effects as well as signal
and background process modeling are symbolized as parameters \lbd and affect the expected event yields.  
For each bin $i$, a Poisson likelihood $P$ is derived from $N^p_i$ and the number of observed
events $N^d_i$. 
The values for \xsec and \mtmc are determined from the maximum $L_\text{max}$ of the global likelihood 
\begin{equation}
L(\xsec,\mtmc,\lbd)=\prod_i 
P\left(N^p_i(\xsec,\mtmc,\lbd),N^d_i\right)
 \cdot \Xi(\lbd) \text{.}
\end{equation}
Here, $\Xi(\lbd)$ represents optional terms that can model prior
knowledge on the systematic uncertainties specific to the
experiment. Alternatively, the fit can be repeated for each individual
systematic variation, leaving only \mtmc and \xsec as free parameters. 

Explicit correlations between \xsec and \mtmc are introduced by the term $\epsilon(\mtmc,\lbd)$.
Hence, the contribution of \mtmc to the total uncertainty on \xsec can be
minimized by reducing the dependence of $\epsilon$ on \mtmc  
or by the strong constraints on \mtmc through $n_i^p$.

\updated{The dependence of the resulting measured cross section on \mtmc has been diminished and absorbed into the uncertainty, 
while the predicted cross section $\xsec^p$ remains a function of \mt. Therefore, \mt is given by the value at which 
the predicted and measured cross sections coincide.} For calculating the uncertainties on $\Delta_m$, correlations between \xsec
and \mtmc need to be accounted for but are known precisely as a result of the
simultaneous fit.

\bigskip
Precise measurements of the inclusive \ttbar cross section are performed in
the dileptonic decay channel by the ATLAS and CMS
collaborations~\cite{Aad:2014kva,CMS-PAS-TOP-13-004}. The uncertainties of
these measurements are below 4\% and the dependence on \mtmc is small. 
In both analyses, \mtpole is extracted assuming $|\Delta_m| \lesssim 1\GeV$,
and assigning a corresponding uncertainty. 
The resulting total precision of \mtpole is about 2\GeV~\cite{CMS-PAS-TOP-13-004}. 
Measurements of \mtmc have been performed in the same \ttbar decay channel
using LHC data at a center-of-mass energy of
$\sqrt{s}=7$~or~8\TeV~\cite{CMS-PAS-TOP-14-014,Aad:2015nba}. The value of
\mtmc is extracted from the normalized  distribution of the lepton and b-jet
invariant mass \mlb. The resulting precision is about 1.3\GeV and 
the dominant uncertainties of both measurements are mostly orthogonal. 
Therefore, combining these analyses, the correlation between the simultaneously determined \xsec and \mtmc will become small.

For illustration, we use the \ttbar production cross section, measured in
Ref.~\cite{Kieseler:2015} at $\sqrt{s}=8\TeV$, 
$\xsec=243.9\, \pm\, 9.3\, \text{pb}$ to determine \mtmsbar and \mtpole for different orders of perturbative QCD. 
The LHC beam-energy uncertainty of 1.72\% is assigned to the predicted cross
section, evaluated with the program HATHOR~\cite{Aliev:2010zk} 
based on calculations of Refs.~\cite{Langenfeld:2009wd,Baernreuther:2012ws,Czakon:2012zr,Czakon:2012pz,Czakon:2013goa}.
The cross section is calculated at LO, NLO, and 
next-to-next-to leading order (NNLO) accuracy with $\alpha_S$ at the Z-boson mass $M_Z$
set to $\alpha_S(M_Z)=0.118\pm 0.001$ and is obtained using the parton distribution
(PDF) set CT14~\cite{Dulat:2015mca} evaluated at NNLO. 
Renormalization and factorization
scales are set to \mtpole or \mtmsbar, respectively, and are varied independently
by a factor of 2 up and down. 
The uncertainties due to variations of the CT14 PDF eigenvectors are scaled to 68\% confidence level. 

\begin{figure}[h!]
\includegraphics[width=0.475\textwidth]{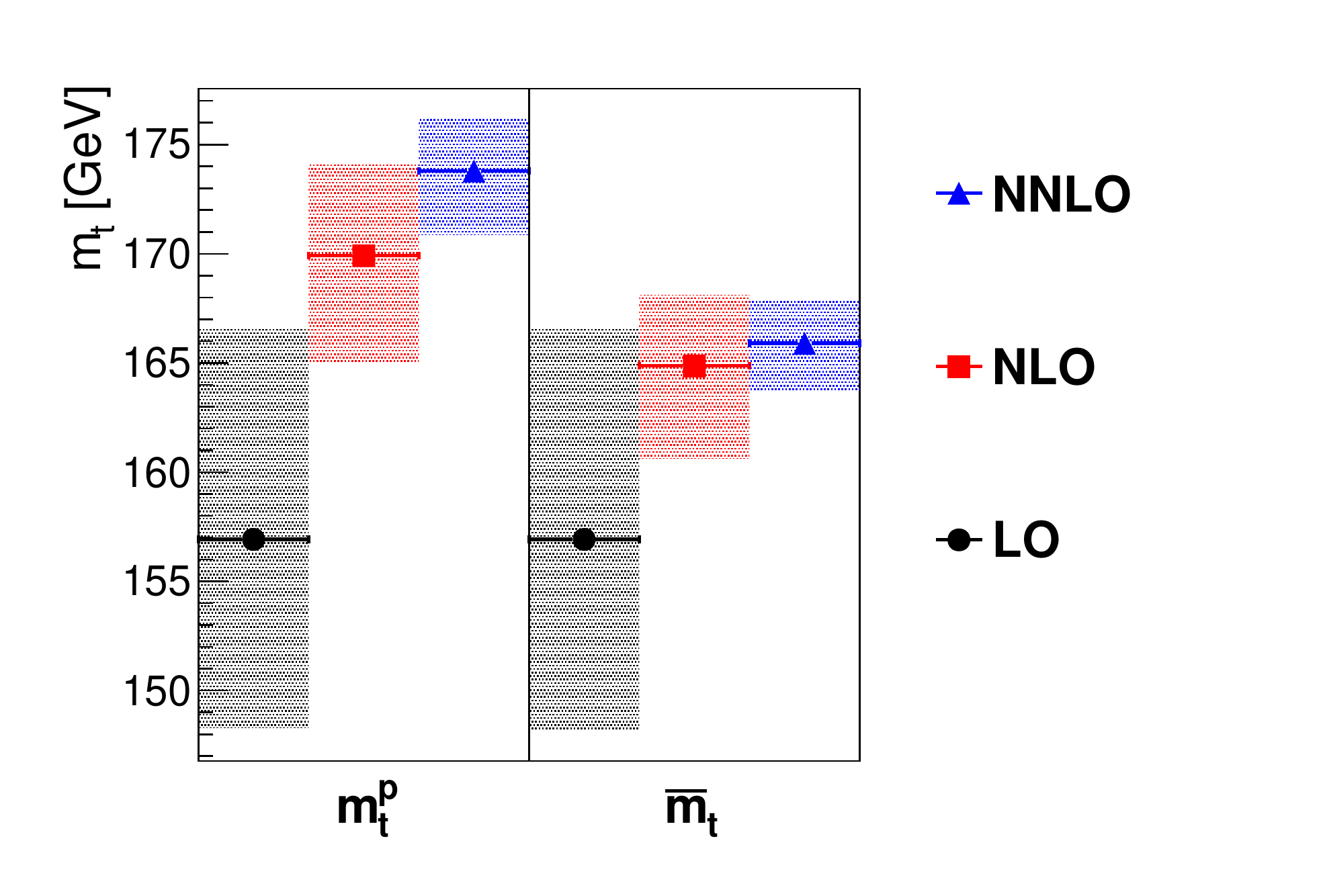}
\caption{Top-quark pole (\mtpole) and \MSbar mass (\mtmsbar) extracted from the inclusive \ttbar production cross section by comparison with its prediction at different orders of perturbative QCD. The hatched areas indicate the total uncertainty on the measured mass values.
\label{fig:convergence}}
\end{figure}

The extraction of \mtpole and \mtmsbar is performed by comparison of predicted
and measured \xsec.  Experimental and theoretical uncertainties are considered
uncorrelated. 
The resulting top-quark mass values are illustrated in Fig.~\ref{fig:convergence}. 
The scheme choice does not play a role at LO. 
When higher orders are considered in the calculation of \xsec,
\mtmsbar exhibits a more rapid convergence than \mtpole. 

\begin{figure}[h!]
\includegraphics[width=0.475\textwidth]{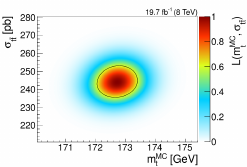}
\caption{Likelihood $L$ for the measured MC mass ($\mtmc$) and \ttbar production cross section ($\xsec_{t\bar{t}}$) at a center-of-mass energy of 8\TeV. The black contour corresponds to the 1 sigma uncertainty~\cite{Kieseler:2015}.
\label{fig:simfit}}
\end{figure}

A detailed experimental analysis employing the method proposed here is
documented in Ref.~\cite{Kieseler:2015}: the fit of \mtmc and \xsec is performed simultaneously at center-of-mass energies of 7 and 8\TeV. As illustrated in Figure~\ref{fig:simfit}, the measured values are mostly uncorrelated.

The obtained cross sections are compared to calculations with NNLO accuracy to determine \mtmsbar. For the extraction of \mtpole, next-to-next-to leading log (NNLL) contributions are also accounted for. 
The measured  \mtmsbar is converted to the pole mass \mtpoleconv 
in perturbation theory with up to four-loop accuracy in QCD~\cite{Marquard:2015qpa}. 
It is well-known that this leads to an additional positive shift of the value of \mtpole, 
the size of which indicates the residual theoretical uncertainty on \mtpole at yet higher orders.
\updated{For example}, using a fixed \mtmsbar as input, the value of \mtpole is approximately 0.5\GeV\!\!(0.2\GeV)
larger if the conversion formula is applied at three(four)-loop instead of
two(three)-loop accuracy, respectively.

The results obtained at $\sqrt{s}=$~7 and 8\TeV for \mtmsbar, \mtpole, and \mtpoleconv are listed in Table~\ref{tab:exmasses} for different PDF sets~\cite{Alekhin:2013nda,Ball:2014uwa,Harland-Lang:2014zoa,Dulat:2015mca}. A strong correlation between the strong coupling constant, \alphas, and the measured top-quark mass can be observed.
\begin{table}[h!] \centering 
\renewcommand{\arraystretch}{1.25}
        \begin{tabular}{ l| c | c | c | c }
       &  $\alphas(M_Z)$ & $\mtmsbar$ [$\GeV$] & $\mtpole$ [$\GeV$]  & $\mtpoleconv$ [$\GeV$] \\ \hline
ABM12    & 0.113  &   $158.4 \pm^{1.2}_{1.9}$&   $166.6 \pm^{1.6}_{1.9}$   & $168.0 \pm^{1.3}_{2.1}$\\
NNPDF3.0 & 0.118  &   $165.2 \pm^{1.1}_{1.7}$&   $174.0 \pm^{1.4}_{1.7}$   & $175.1 \pm^{1.2}_{1.9}$\\
MMHT2014 & 0.118  &   $165.4 \pm^{1.1}_{1.9}$&   $174.3 \pm^{1.4}_{1.8}$   & $175.3 \pm^{1.3}_{2.1}$\\
CT14     & 0.118  &   $165.5 \pm^{1.5}_{2.0}$&   $174.4 \pm^{1.8}_{2.0}$   & $175.4 \pm^{1.7}_{2.2}$\\
\end{tabular}
\caption{Measured \MSbar ($\mtmsbar$), pole ($\mtpole$), and pole mass from conversion ($\mtpoleconv$) for different PDF sets and values for the strong coupling constant, \alphas, evaluated at the Z-boson mass, $M_Z$~\cite{Kieseler:2015}.
 \label{tab:exmasses}}
\end{table}
All extracted values for \mt are used to
calibrate the \mtmc parameter, which is non-universal and, in principle,
depends on the subtleties of its implementation in the MC simulation. 
In Ref.~\cite{Kieseler:2015},  
$\bar{\Delta}_m = \mtmsbar-\mtmc$, $\Delta_m^p = \mtpole-\mtmc$, and $\Delta_m^{p,c} = \mtpoleconv-\mtmc$ are calculated 
for \mtmc as implemented in MadGraph5~\cite{Alwall:2011uj} interfaced with
Pythia6~\cite{Sjostrand:2006za} \updated{using the tune Z2$^*$~\cite{Chatrchyan:2013gfi} 
and top-quark decays simulated with MadSpin2~\cite{Frixione:2007zp}.}
The results are listed in Table~\ref{tab:calib}. A precision of about 2\GeV is achieved.

\begin{table}[h] \centering 
\renewcommand{\arraystretch}{1.25}
        \begin{tabular}{ l| c | c | c } 
          & $\bar{\Delta}_m$ [$\GeV$] & $\Delta_m^p$ [$\GeV$]    & $\Delta_m^{p,c}$ [\GeV]  \\ \hline
ABM12     &   $-14.3 \pm^{1.4}_{2.0}$&   $-6.1 \pm^{1.7}_{2.0}$  &   $-4.7 \pm^{1.5}_{2.2}$ \\
NNPDF3.0  &   $-7.6 \pm^{1.3}_{1.9}$  &   $1.3 \pm^{1.6}_{1.9}$  &   $2.4 \pm^{1.5}_{2.0}$ \\
MMHT2014 &   $-7.3 \pm^{1.3}_{2.1}$  &   $1.5 \pm^{1.6}_{2.0}$   &   $2.6 \pm^{1.5}_{2.2}$ \\
CT14     &   $-7.2 \pm^{1.7}_{2.1}$  &   $1.6 \pm^{1.9}_{2.1}$   &   $2.7 \pm^{1.8}_{2.3}$ \\
\end{tabular}
\caption{Difference between the top quark mass in well-defined schemes and the top-quark MC mass for different PDF sets. The MC mass is compared to the \MSbar mass ($\bar{\Delta}_m$), pole mass ($\Delta_m^p$), and the pole mass from conversion ($\Delta_m^{p,c}$)~\cite{Kieseler:2015}.
 \label{tab:calib}}
\end{table}

\section{Calibration with differential cross sections}

An extension of the method to differential cross sections used for the
determination of \mt can provide a larger sensitivity and, possibly, a further
reduction of systematic uncertainties.  
In the following, a differential production cross section for the signal
process as a function of an observable $x$ is considered and employed to
determine \mt. 
The approach used for \xsec is applied to each bin of this differential cross section.
For this purpose, the  efficiency $\epsilon$ is replaced by a matrix $M$
describing the detector response to the predicted cross section $\xsec^\text{MC}_k$ 
in bin $k$ of the distribution in terms of $x$, defined
by: 
\begin{equation}
\label{eq:fold}
{N}^{ps}_j = \mathcal{L} \cdot \sum_k M_{jk} {\xsec^\text{MC}_k} \text{,}
\end{equation}
with ${N}^{ps}_j$ being the predicted number of reconstructed and selected
signal events in bin $j$ of the reconstructed distribution. The response
matrix is derived from MC simulation and therefore depends on \lbd as well as
on \mtmc~\footnote{A more complete discussion of the response matrix can be found in
Ref.~\cite{CMS-PAS-TOP-14-014}.}.

Each bin $j$ of the reconstructed distribution is considered as a category. In
each category,  a second observable $y$ is defined, sensitive to \mtmc. The
shape of this observable is used to constrain \mtmc, while the total number of
signal events in each category corresponds to ${N}^{ps}_j$, and hence can be
used to derive the differential cross section. 
 The number of predicted events, $N^p_{ij}$, in bin $i$ of the observable $y$ is given as:
\begin{equation}
N^p_{ij}= \mathcal{L} \cdot \sum_k M_{jk}(\mtmc,\lbd)\ {\xsec^\text{MC}_k} \cdot n_{ij}^p(\mtmc,\lbd) 
\\ +N^{bg}_{ij}(\lbd) \text{,}
\end{equation}
with $n_{ij}^p$  being the fraction of predicted signal events in bin $i$ with
respect to ${N}^{ps}_j$ and $N^{bg}_{ij}$ the contribution from background
processes. 

 By comparison with the number of observed events $N_{ij}^d$ in each
 category $j$ and bin $i$, and  considering $\xsec^\text{MC}_k\rightarrow
 \xsec_k$ as free parameters a fit can be performed maximizing the likelihood: 
\begin{equation} 
 L(\xsec_0,...,\xsec_k,\mtmc,\lbd)=\prod_i \prod_j
P\left(N^p_{ij},N^d_{ij}\right)
 \cdot \Xi(\lbd) \text{.}
\end{equation}
This unfolding problem can be ill-posed and regularization techniques might need to be applied. 
A well-suited regularization condition is provided, for instance, by the aim to determine \mt by
comparison of $\xsec_k$ with its prediction $\xsec^p_k(\mt)$ as a function
of \mt. Replacing $\xsec_k$ with this prediction corresponds to the folding
approach used in Ref.~\cite{CMS-PAS-TOP-14-014} and reduces the number of free
parameters significantly, such that the likelihood becomes: 
\begin{equation} 
 L(\mt,\mtmc,\lbd,\vec{\kappa})=\prod_i \prod_j
P\left(N^p_{ij},N^d_{ij}\right)
 \cdot \Xi(\lbd,\vec{\kappa}) \text{,}
\end{equation}
with $\Xi(\lbd,\vec{\kappa})$ representing optional nuisance terms and
$\vec{\kappa}$ being theoretical uncertainties on the predicted
$\xsec^p_k(\mt)$. Both, \lbd and $\vec{\kappa}$ can be incorporated as
nuisance terms in $\Xi$ or can be evaluated individually. In the latter case,
$L$ depends on \mt and \mtmc, only.  
A maximization of $L$ directly returns the relation between these parameters as well as their correlations.
The correlations are mainly incorporated through the response matrix
$M$. Therefore, the event selection and the observable $x$ should be chosen
such, that the dependence of $M$ on \mtmc is minimized and the sensitivity of
$y$ on \mtmc becomes maximal.  

For the optimization of the result, also the correlation between the observables $x$ and $y$ should be small.
A possible choice for $x$ would be the differential \ttbar production cross
section as a function of the top-quark transverse momentum predicted up to NNLO accuracy~\cite{Czakon:2015xx}. The dependence of
this observable on \mtpole and \mtmsbar can be studied at approximate NNLO
 with programs publicly available~\cite{Guzzi:2014wia}. 
This distribution, describing the production dynamics, can be combined with an
observable based on the kinematics of the decay products such as \mlb in the
dileptonic decay channel or the invariant mass of the 3 jets that originate
from the top-quark decay $t\rightarrow W b \rightarrow b q\bar{q}$ in the semileptonic channel. 

The additional sensitivity of the differential cross sections to \mt can
result in uncertainties below 2\GeV on \mt and $\Delta_m$, starting to
challenge the measurements of \mtmc in precision and improving the
understanding of this parameter. 
Moreover, determinations of the running of $\mtmsbar(\mu)$ at varying scales $\mu$ 
as well as simultaneous extractions of the strong coupling $\alpha_S$ and \mt become possible.

\section{Conclusion}


\updated{The simultaneous determination of \mtmc and of differential or inclusive production cross sections 
of processes sensitive to the top-quark mass \mt allows for subsequent extraction of \mt in a well-defined 
renormalization scheme. This method solves the longstanding problem of the calibration of the top-quark 
Monte Carlo mass \mtmc and, in addition, allows for a consistent quantification of the difference $\Delta_m=\mt-\mtmc$ 
for the particular MC tools used in the analysis and within the uncertainties of the measurement.} 

The extraction of \mt is preferably performed in a scheme, where the perturbative expansion of 
the theory prediction for the respective cross section displays fast apparent convergence.
\updated{For inclusive cross section, 
this applies to} short-distance masses and favors an experimental determination 
of a running top-quark mass \mtmsbar over the pole mass \mtpole. 
The extracted \mtmsbar is more precise than \mtpole obtained at the same order of
perturbation theory and additional higher-order corrections result in
smaller corrections to \mtmsbar than \mtpole. 
The latter can always be obtained up to four-loop accuracy in QCD.

With the current precision of the inclusive top-quark cross-section and mass
measurements an uncertainty on $\Delta_m$ of approximately 2\GeV can be achieved.
Dedicated analyses based on differential cross sections seem to be a promising
approach to further decrease this uncertainty and to measure theoretically
well-defined mass parameters independently of the interpretation of the top-quark MC mass to a high precision.

\smallskip
{\bf{Acknowledgments}}\\
We would like to thank Olaf Behnke for useful discussions. 


%

\end{document}